# Quantum-Safe Web Service Architecture Using Time-Based One-Time Passwords

Abel C. H. Chen
*Information & Communications Security Laboratory,*
*Chunghwa Telecom Laboratories*
Taoyuan, Taiwan
Email: chchen.scholar@gmail.com; ORCID: 0000-0003-3628-3033

***Abstract*—One-Time Passwords (OTPs) have become a common option for multi-factor authentication in several applications. For instance, during website login processes, OTPs are often used in conjunction with traditional text-based usernames and passwords to verify whether the access request originates from a legitimate human user rather than an automated agent. However, in scenarios involving automated connections and system-to-system interoperability, Time-Based One-Time Passwords (TOTPs) may be required to establish secure connections and access Web Services (WSs). Therefore, this study focuses on exploring the development of a quantum-safe web service architecture. The proposed approach achieves transmission security management by implementing Transport Layer Security (TLS) and HyperText Transfer Protocol Secure (HTTPS) based on Post-Quantum Cryptography (PQC). Furthermore, web service security management is realized through the construction of keyed-Hash Message Authentication Code (HMAC)-driven TOTPs. Within the experimental environment, this study evaluates and compares the computational performance of the Secure Hash Algorithm-2 (SHA-2), SHA-3, Ascon-Hash256, and SM3. The required computation time under different hardware resource conditions is analyzed for future web service deployment.**

## I. INTRODUCTION

In recent years, the number of malicious activities on the Internet has continued to increase [1], [2], posing potential threats to users' financial assets [3] and personal privacy [4]. Consequently, many websites have deployed One-Time Password (OTP) mechanisms for authentication purposes [5]. However, in scenarios involving automated connections and system-to-system interoperability, secure connections and access to Web Services (WSs) require the use of Time-Based One-Time Passwords (TOTPs) [6], [7]. Currently, the generation of TOTPs is commonly based on the combination of a keyed-Hash Message Authentication Code (HMAC) [8], [9] and a timestamp. Since the security of TOTP mechanisms relies on the underlying hash functions, the pseudo-random numbers generated by quantum-safe hash functions (e.g. Secure Hash Algorithm-384 (SHA-384)) could be considered for TOTP generation [10]. For instance, the authentication mechanisms used during login in the National Institute of Standards and Technology (NIST) Automated Cryptographic Validation Test System (ACVTS) [11] and the Automated Cryptographic Validation Protocol (ACVP) [12] adopt TOTPs.

At present, many websites generate pseudo-random numbers using the SHA-2 in conjunction with HMAC and subsequently derive TOTPs from these values [11]. However, limited comparative analysis has been conducted with other hash functions, such as SHA-3 [13], Ascon-Hash256 [14], and SM3 [15]. Therefore, this study focuses on the development of a quantum-safe web service architecture by integrating Post-Quantum Cryptography (PQC) into Transport Layer Security (TLS) [16] and HyperText Transfer Protocol Secure (HTTPS) [17] to enhance transmission security. Furthermore, this study designs an HMAC-driven TOTP mechanism in which different hash functions are incorporated into the HMAC construction, and a comparative evaluation of computational performance is conducted based on measured computation time.

The major contributions of this study are summarized as follows.

- This study proposes a quantum-safe web service architecture based on PQC, in which both web service servers and client devices independently possess PQC-based X.509 certificates [18]. PQC mechanisms are integrated into the TLS protocol to enable secure communication and mutual authentication.
- This study considers HMAC constructions based on different hash functions and designs an HMAC-driven TOTP mechanism. The proposed mechanism generates OTPs using a pre-shared seed and a timestamp for each authentication instance.
- This study implements and evaluates the computation time of HMAC-driven TOTP generation under different hash functions (e.g., SHA-512, SHA3-512, SHA-384, SHA3-384, SHA-256, SHA3-256, Ascon-Hash256, and SM3). Experimental implementations are conducted on both a general-purpose personal computer and a Raspberry Pi 4 platform.

The remainder of this paper is organized as follows. Section II presents the research background and reviews the core concepts and current developments of the related technologies. Section III describes the proposed architecture, addressing both transmission security management and web service security management. Section IV presents the implementation details and comparative experimental results. Finally, Section V concludes the paper and discusses future research directions.

## II. BACKGROUND AND RELATED WORK

This section presents the main ideas of HMAC, TOTP, and TLS.

### *A. Keyed-Hash Message Authentication Code*

The NIST published the FIPS 198-1 standard [8], which defines the steps of HMAC, as illustrated in Fig. 1. The input parameters of HMAC consist of a secret key $k$ and a message $Msg$. The computation first applies the inputs to the $Inner(k, Msg)$ function, in which the key $k$ is combined with a padding value $P_b$ using a bitwise exclusive OR (XOR) operation. The resulting value is then concatenated with *Msg* and processed by the underlying hash function. Assuming a block size of $b$ bytes, the padding value $P_b$ is defined as a sequence of $b$ bytes with each byte set to 0x36, while the padding value $P_b'$ consists of $b$ bytes with each byte set to 0x5C. Finally, the output of *Outer*($k$) is concatenated with the result of *Inner*($k$, *Msg*) and passed through the hash function to produce the final HMAC value, denoted as *HMAC*($k$, *Msg*).

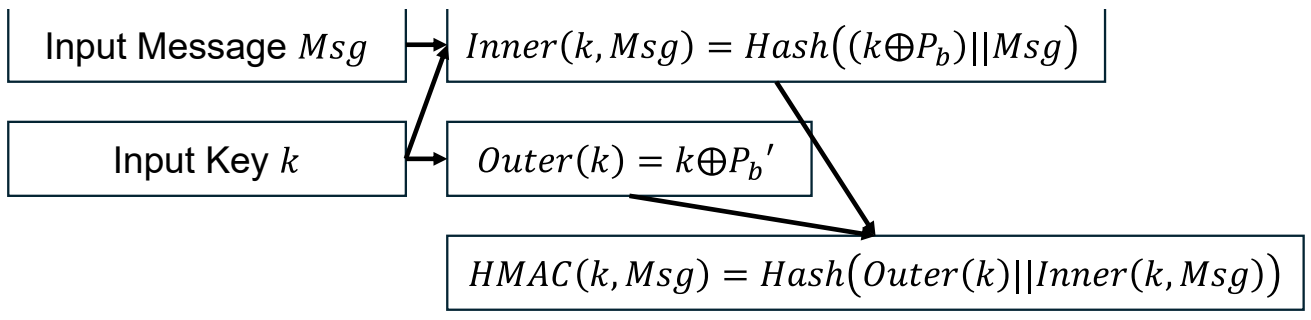

Fig. 1. The detailed steps of HMAC.

## B. Time-Based One-Time Passwords,

The login process of the NIST ACVTS requires the use of TOTPs. Prior to authentication, a pre-shared seed is established between the web service server and the client device. During subsequent login attempts, the pre-shared seed is used as the secret key $k$, and the current timestamp is used as the message *Msg*. These inputs are processed through a HMAC to generate the TOTP, as illustrated in Fig. 2.

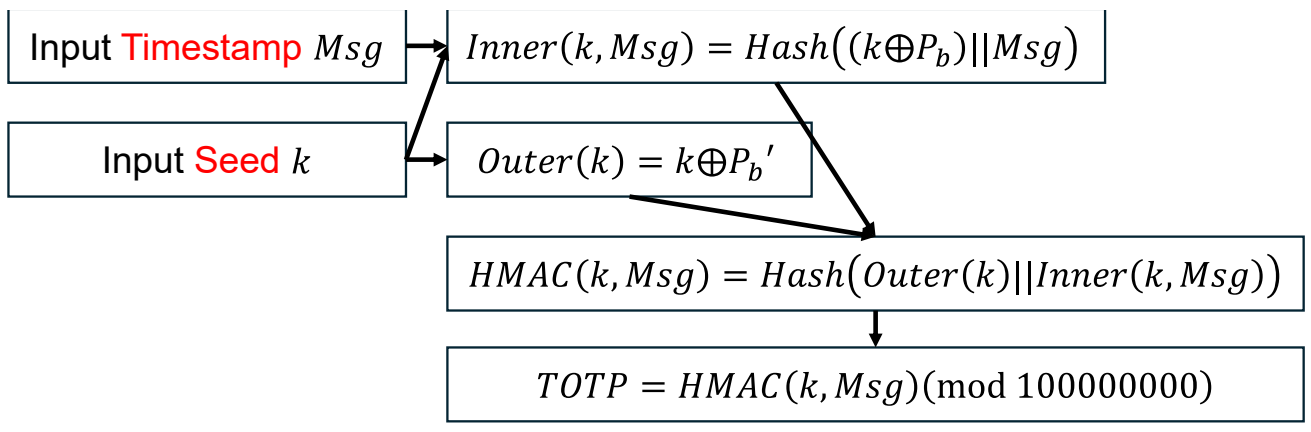

Fig. 2. The detailed steps of HMAC-driven TOTP.

The timestamp is typically divided into fixed time steps of 30 seconds, such that the timestamp is divided by 30 seconds and the integer value is used. Under the assumption that both the web service server and the client device possess the same pre-shared seed and are synchronized within the same 30-second time window, identical TOTPs can be independently generated. Authentication success is determined by comparing the generated TOTPs for consistency.

In addition, to constrain the TOTP output to a fixed number of decimal digits, a modulo operation is applied. For instance, an 8-digit decimal TOTP is obtained by computing the value modulo 100,000,000, whereas a 10-digit decimal TOTP is obtained by computing the value modulo 10,000,000,000.

## C. Transport Layer Security

TLS version 1.3 [16] is one of the mainstream protocols for establishing secure communication at the transport layer. It negotiates cryptographic algorithms and session keys between communicating parties through the exchange of ClientHello and ServerHello messages. This section uses the NIST ACVTS as an illustrative example. In this scenario, the client device has previously obtained a valid certificate, and the public key contained in the certificate is based on RSA cryptography.

The TLS 1.3 handshake process is illustrated in Fig. 3. First, the client device generates a ClientHello message containing the cipher_suite, key_share, and signature_algorithms fields. The client device generates an ephemeral secp384r1 key pair using the NIST P-384 elliptic curve parameters. The cipher_suite is set to TLS_ECDHE_RSA_WITH_AES_256_GCM_SHA384, and the key_share field carries the ephemeral secp384r1 public key, indicating that Elliptic Curve Diffie–Hellman Ephemeral (ECDHE) key exchange based on NIST P-384 is used, with HMAC-SHA384 employed for message authentication. The signature_algorithms field is set to rsa_pkcs1_sha256, specifying that the server will subsequently generate signatures using this algorithm.

Upon receiving the ClientHello message, the web service server generates a ServerHello message, which includes the cipher_suite, key_share, CertificateRequest, Certificate, CertificateVerify, Finished, and Application Data messages. The server also generates an ephemeral secp384r1 key pair using the NIST P-384 parameters. The cipher_suite is set to TLS_ECDHE_RSA_WITH_AES_256_GCM_SHA384, and the key_share field contains the server's ephemeral secp384r1 public key. Both parties then derive a shared secret using their own private keys and the peer's public key. To achieve mutual authentication, the server issues a CertificateRequest message. The Certificate message carries the server's certificate, while the CertificateVerify message includes a signature that can be verified using the public key contained in the certificate. Finally, the Finished message contains a message authentication code (MAC) generated using HMAC-SHA384, which is used to verify the integrity of the handshake messages.

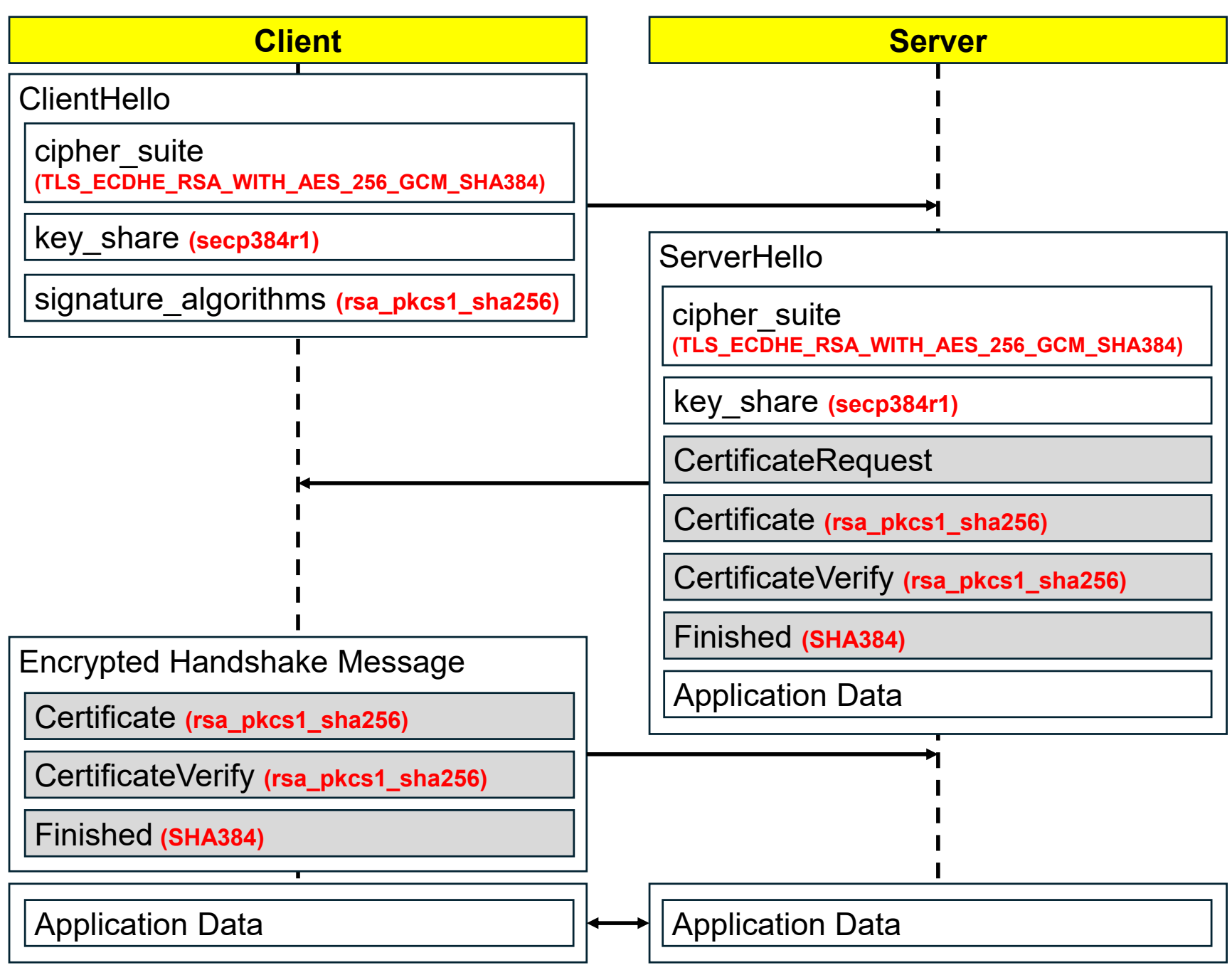

Fig. 3. An example of transport layer security.

After receiving the ServerHello message, the client device first derives the shared secret and subsequently generates an AES encryption key through a key derivation function. The client device then constructs the Encrypted Handshake Messages, which include the Certificate, CertificateVerify, and Finished messages. The Certificate message carries the client device's certificate, while the CertificateVerify message contains a signature that can be verified using the public key embedded in the certificate. Finally, the Finished message includes a MAC generated using HMAC-SHA384 to confirm the integrity of the handshake messages, and the handshake messages are encrypted using the derived AES key.

## III. Quantum-Safe Web Service Architecture

This section illustrates the main ideas of the proposed quantum-safe web service architecture including transmission security management and web service security magagement.

### *A. Transmission Security Management*

To establish a quantum-safe web service architecture, this study recommends that client devices submit certificate signing requests to a certificate authority (CA) through out-of-band communication channels (e.g., e-mail) prior to service access. The CA then issues verifiable certificates [18], in which the public keys are based on the Module-Lattice-Based Digital Signature Algorithm (ML-DSA).

The proposed quantum-safe TLS handshake process is illustrated in Fig. 4. First, the client device generates a ClientHello message containing the cipher_suite, key_share, and signature_algorithms fields. The client device generates an ephemeral X25519 key pair based on elliptic curve cryptography (ECC) and an ephemeral ML-KEM-768 key pair based on the Module-Lattice-Based Key-Encapsulation Mechanism (ML-KEM). The cipher_suite is set to TLS_CHACHA20_POLY1305_SHA256, and the key_share field is populated with the concatenation of the ephemeral X25519 public key and the ML-KEM-768 public key, indicating a hybrid key exchange that combines ECC and PQC. HMAC-SHA256 is employed for message authentication. The signature_algorithms field is set to id-ml-dsa-44, specifying that the web service server will generate signatures using this algorithm.

Upon receiving the ClientHello message, the web service server generates a ServerHello message containing the cipher_suite, key_share, CertificateRequest, Certificate, CertificateVerify, Finished, and Application Data messages. The server generates an ephemeral X25519 key pair and produces an ML-KEM-768 key encapsulation using the client's ML-KEM-768 public key. The cipher_suite is set to TLS_CHACHA20_POLY1305_SHA256, and the key_share field contains the concatenation of the server's ephemeral X25519 public key and the ML-KEM-768 key encapsulation value. Both parties subsequently derive a shared secret. To achieve mutual authentication, the server issues a CertificateRequest message. The Certificate message carries the server's ML-DSA-based certificate, while the CertificateVerify message includes a signature that can be verified using the ML-DSA public key contained in the certificate. Finally, the Finished message carries a MAC generated using HMAC-SHA256 to confirm the integrity of the handshake messages.

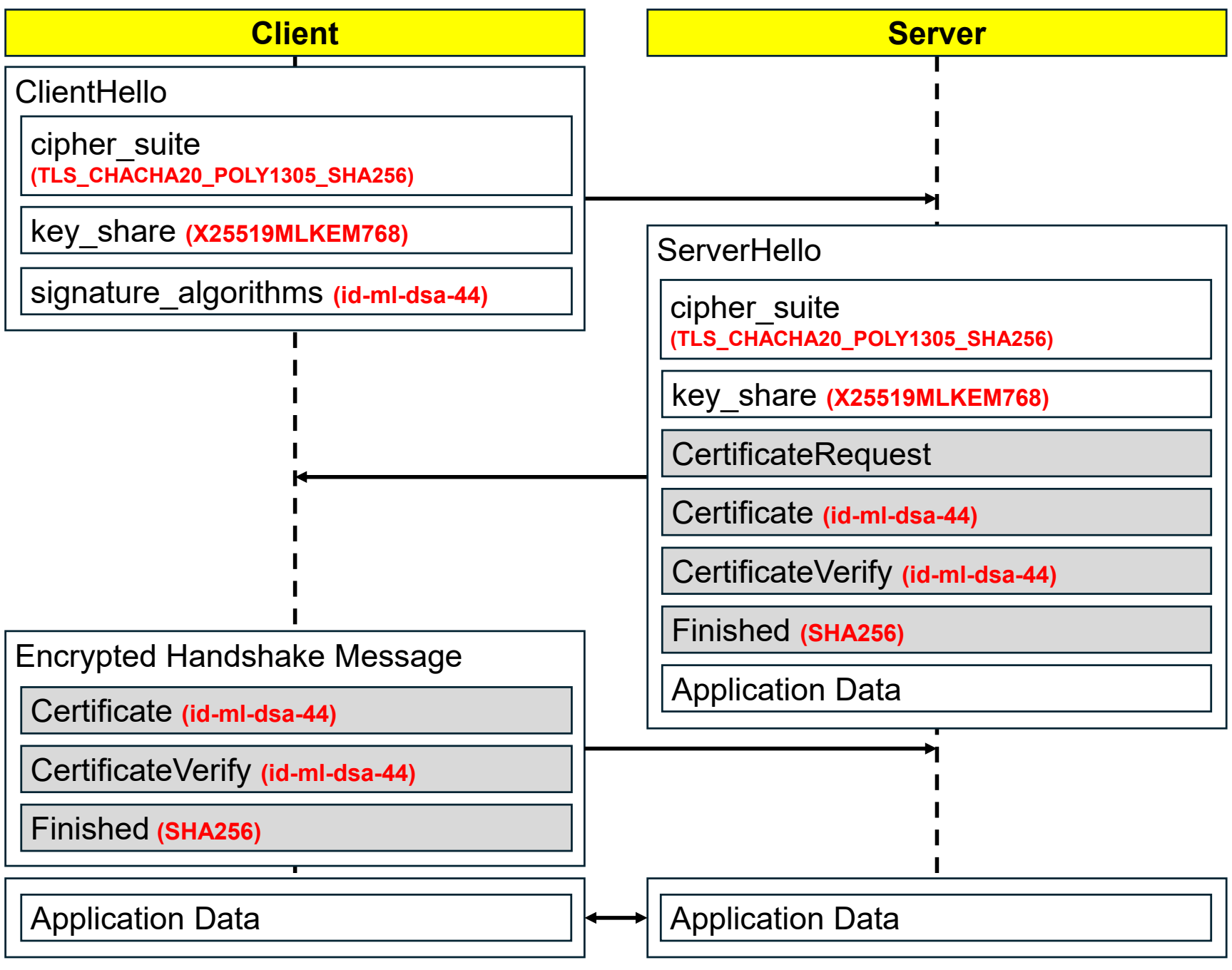

Fig. 4. An example of the proposed quantum-safe transport layer security.

After receiving the ServerHello message, the client device first derives the shared secret and subsequently generates a ChaCha20-Poly1305 encryption key through a key derivation function (KDF). The client device then constructs the Encrypted Handshake Messages, which include the Certificate, CertificateVerify, and Finished messages. The Certificate message carries the client device's ML-DSA-based certificate, while the CertificateVerify message contains a digital signature that can be verified using the ML-DSA public key embedded in the certificate. Finally, the Finished message includes a MAC generated using HMAC-SHA256 to confirm the integrity of the handshake messages, and the handshake messages are encrypted using the derived ChaCha20-Poly1305 key.

After the communicating parties have successfully established the encryption keys, the derived keys are subsequently used to generate Encrypted Application Data. The payloads of the HTTPS can then be encapsulated within the encrypted application data, thereby enabling secure HTTPS communication.

### B. *Web Service Security Management*

Although TLS can provide mutual authentication and encrypted communication, web service providers still require mechanisms to manage web service requesters. Therefore, a web service security management framework is necessary to allow client devices to log in, while enabling the web service server to verify the identity and access privileges of the service requesters.

For authentication, this study proposes a HMAC-driven TOTP mechanism driven. It is recommended that client devices first submit service access requests to the web service server through out-of-band communication channels (e.g., e-mail). Upon confirmation, the server provides a pre-shared seed to the client device, which can subsequently generate TOTPs following the method described in Section II.B.

To improve the computational efficiency of HMAC-driven TOTPs, this study investigates the use of different hash functions, including SHA-512, SHA3-512, SHA-384, SHA3-384, SHA-256, SHA3-256, Ascon-Hash256, and SM3. Taking Ascon-Hash256 as an example, the hash function used in the computation of *Inner*(*k*, *Msg*) is replaced with Ascon-Hash256, and the hash function used to compute *HMAC*(*k*, *Msg*) is also replaced with Ascon-Hash256. The resulting HMAC value is then reduced modulo 100,000,000 to produce an 8-digit decimal TOTP, as illustrated in Fig. 5.

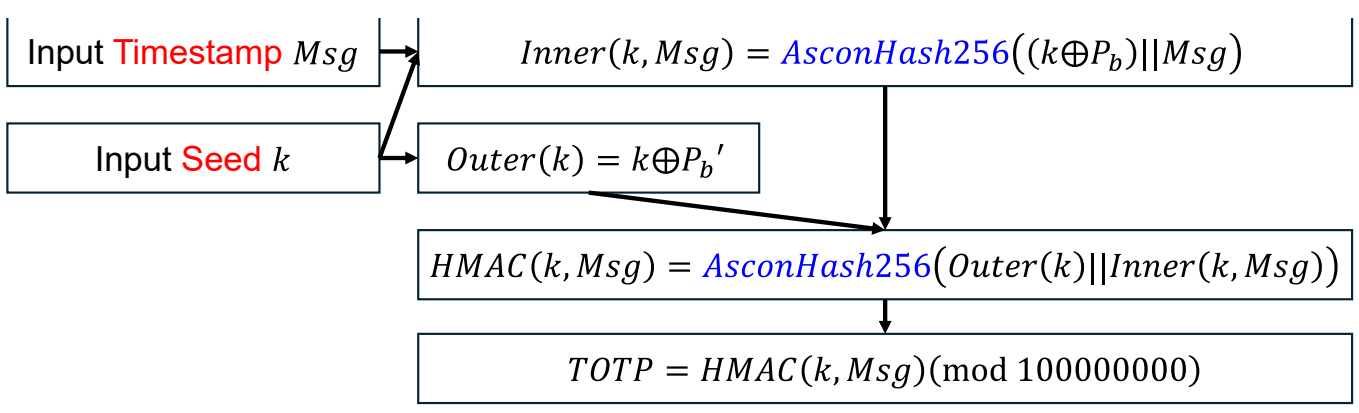

Fig. 5. The detailed steps of the proposed HMAC-driven TOTP.

## IV. Experimental Environment and Experimental Result Discussions

To validate the proposed HMAC-driven TOTP mechanism, the experimental setup and results are described in the following subsections.

### A. *Experimental Environment*

The proposed HMAC-driven TOTP mechanism was implemented on both a general-purpose personal computer (ASUS ExpertBook) and a Raspberry Pi 4. Multiple hash functions were evaluated, including SHA-512, SHA3-512, SHA-384, SHA3-384, SHA-256, SHA3-256, Ascon-Hash256, and SM3. The software environment consisted of JDK 18.0.2.1 and BouncyCastle 1.80. For each hash function, 5,000 TOTP instances were generated, and the computation time required for each generation was recorded.

### B. *Practical Experimental Results and Discussions*

The computation times of the proposed HMAC-driven TOTP mechanism on a general-purpose personal computer and a Raspberry Pi 4 are presented in Fig. 6 and Fig. 7, respectively. The experimental results indicate that, although the absolute computation times differ between the two platforms, the relative distributions and trends are consistent. The results further show that the SHA-2 family exhibits shorter computation times compared to the SHA-3 family, indicating higher computational efficiency for SHA-2 algorithms.

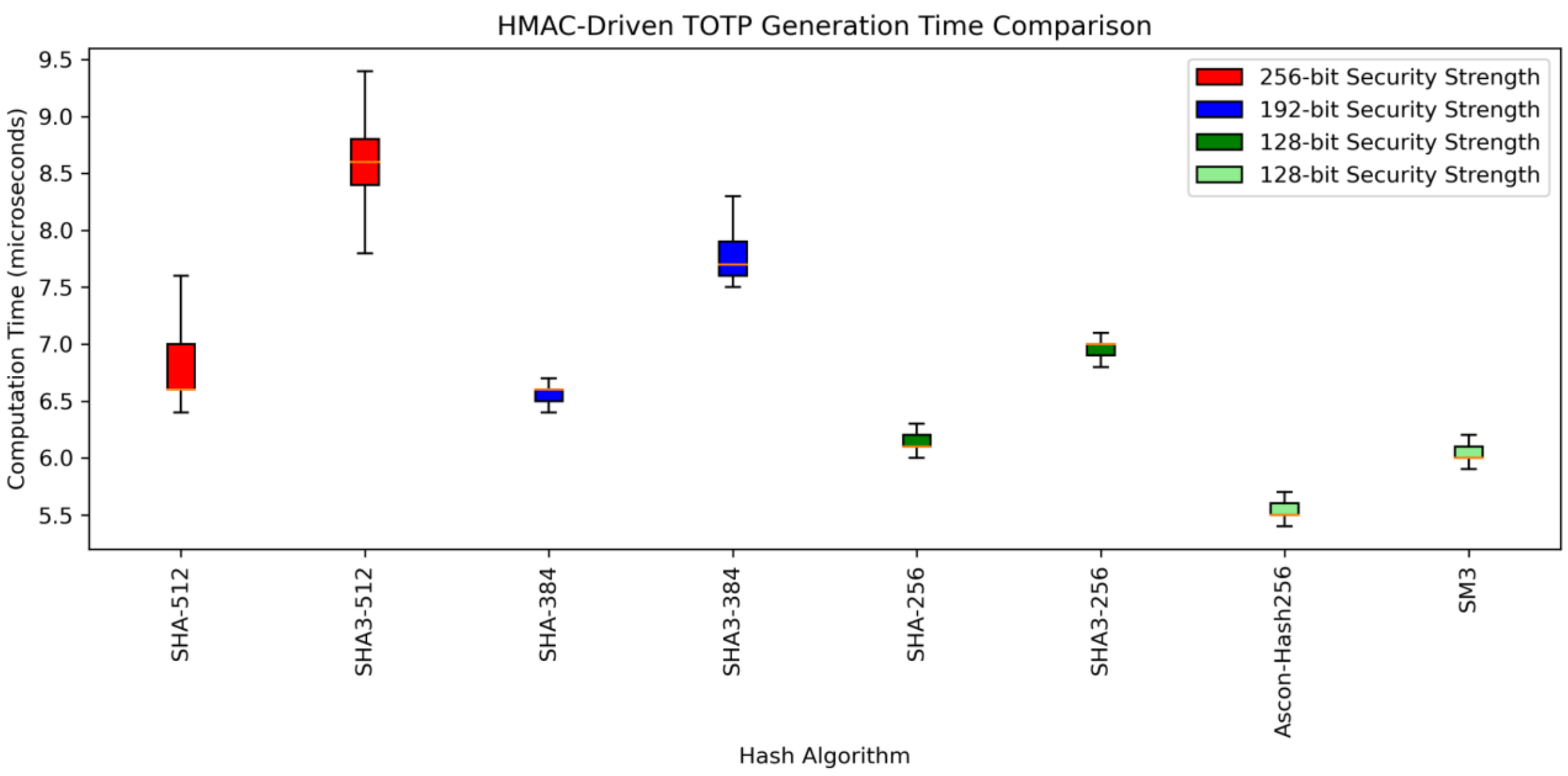

Fig. 6. HMAC-driven TOTP generation time on a personal computer.

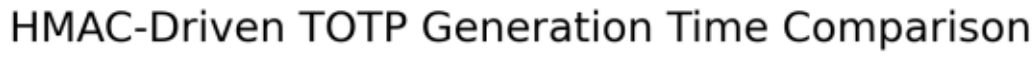

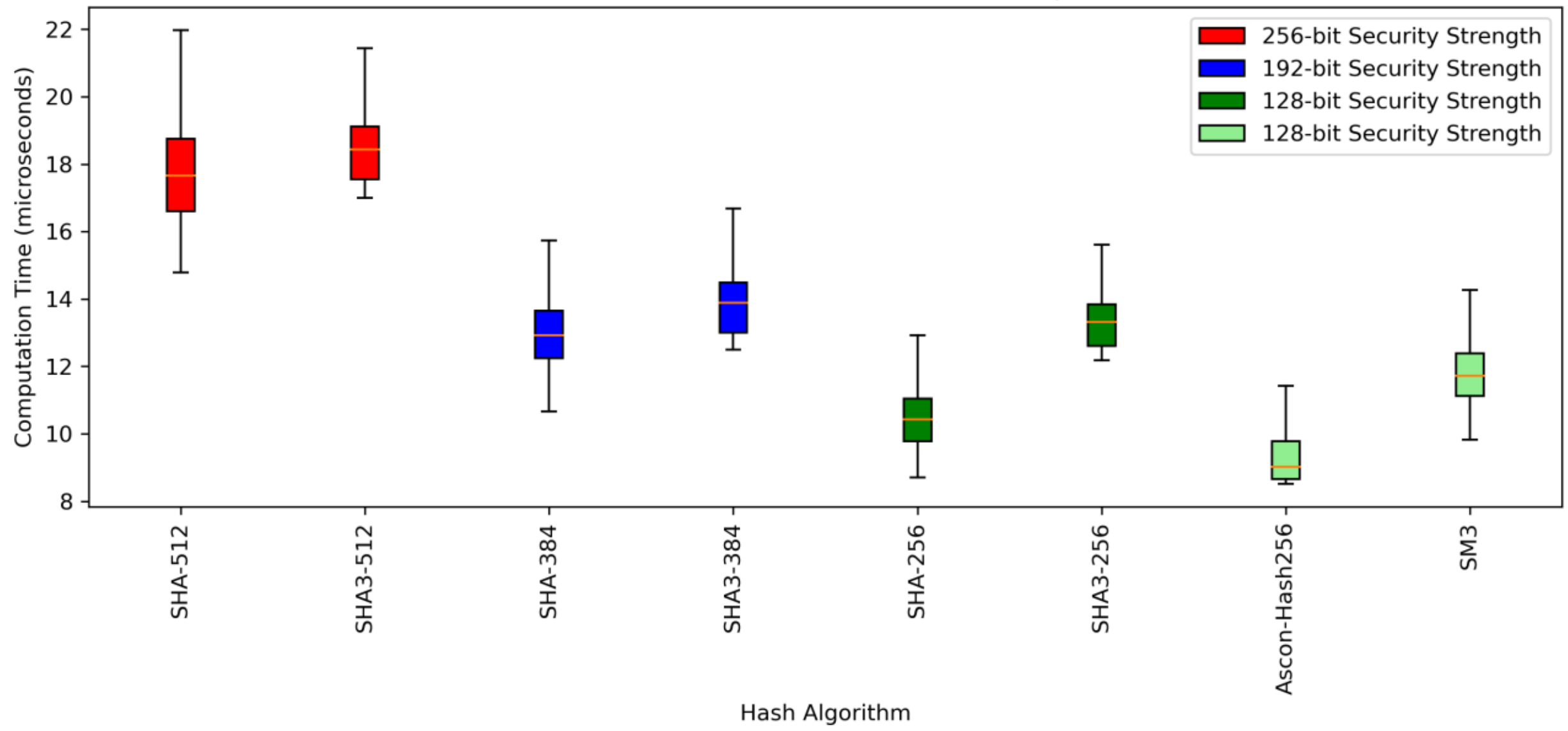

Fig. 7. HMAC-driven TOTP generation time on a Raspberry Pi 4.

Furthermore, when comparing hash functions at equivalent security levels, Ascon-Hash256 demonstrates higher computational efficiency than SHA-256, SHA3-256, and SM3. At this security level, the computation times of SHA-256 and SM3 are comparable, whereas SHA3-256 requires significantly more time than the other methods. These observations suggest that the adoption of HMAC-driven TOTPs using Ascon-Hash256 may improve computational efficiency in future implementations.

## V. Conclusions and Future Work

This study proposes a quantum-safe web service architecture based on PQC, encompassing both transmission security management and web service security management. For transmission security management, a quantum-safe TLS protocol is implemented in conjunction with PQC X.509 certificates, providing quantum-resistant mutual authentication. For web service security management, a HMAC-driven TOTP mechanism is designed, and the computational efficiency of different hash functions (i.e., SHA-512, SHA3-512, SHA-384, SHA3-384, SHA-256, SHA3-256, Ascon-Hash256, and SM3) is evaluated. The experimental results indicate that the HMAC-driven TOTP using Ascon-Hash256 achieves the shortest computation time, suggesting that Ascon-Hash256 may be preferred for future deployment.

Future work may involve deploying the proposed approach across various web service applications and integrating a large number of heterogeneous devices to evaluate both security and computational efficiency under real-world conditions.

## Acknowledgement

This study primarily referenced the NIST ACVTS and the ACVP. The authors gratefully acknowledge NIST for providing access to the demonstration version of the ACVTS, which enabled the successful execution of this research.

The authors also wish to express their appreciation to Mike Ounsworth and John Gray for their support, which facilitated the participation of the PQC X.509 certificates developed in this study in the Internet Engineering Task Force (IETF) Hackathon interoperability testing, successfully passing evaluations conducted by international experts.